# Two – Dimensional Poincare Maps constructed through Ginzburg- Landau (G-L) Theory of critical phenomena in Physics.

**Yiannis Contoyiannis[(1)] and Myron Kampitakis[(2)]**

(1) Department of Electric-Electronics Engineering, West Attica University, 250 Thivon and P. Ralli, Aigaleo, Athens GR-12244, Greece (email: yiaconto@uniwa.gr)
(2) Hellenic Electricity Distribution Network Operator SA, Network Major Installations Department, 72 Athinon Ave., N.Faliro GR-18547, Greece (email: m.kampitakis@deddie.gr)

**Abstract : Based on the saddle point approximation in G-L theory of the critical phenomena we construct two-dimensional Poincare maps which describe the symmetry breaking (SB) and the tricritical crossover phenomenon in Physics. The phase space diagrams of these maps are in agreement with the theoretical predictions. A correction in these maps close to the critical point for small values of the order parameter is attempted. Finally we demonstrate that numerical experiments verify the correctness of these maps.**

**Keywords :** Critical phenomena, Symmetry breaking, Z(N) models , Tricritical crossover, Poincare maps, phase-space.

# 1. Introduction

There are many physical systems whose dynamics reduces to one-or-two dimensional Poincare maps. Hence, a system with many degrees of freedom can be described using the low dimensional Poincare maps. This can be deduced starting from the differential ( or Integrodifferential ) equations for the constituents of the systems or macroscopic variables for which we developed a reduction process. Such techniques have been presented in the past [***H.G. Schuster 1998***, ***E.Ott 1993]***. The advantages of creating such maps from physical, chemical, biological, economic, social, systems are essential and have been extensively reported in the literature. In our opinion, the most important are (a) the numerical processing and (b) the emergence of unknown properties of the systems which appear in phase space. A space that has such information is the phase space, which is the corresponding space of the Statistical Mechanics, with its physical quantity and its derivatives. In this work we will try to produce two dimensional Poincare maps (a) from the G-L mean Field Approximation of the second order phase transition and (b) from tricritical crossover phenomena in phase transitions. We will check whether this system of maps is consistent with the theory of critical phenomena in the level of phase space. Similar works from G-L theory have been presented in the past. These works were based on the G-L Hamiltonian and focused on the similarities between critical phenomena and chaotic Dynamics [***H.G. Schuster. 1998***]. Nevertheless, in our work we use the saddle point approximation on G-L Hamiltonian which better describes the critical state [***N. Antoniou et al 1998, 2000***]. This approximation gives the Euler-Lagrange (E-L) equation from which we produce the Poincare maps. A verification for the correctness of these maps is presented through two numerical experiments from the field of Z(N) spin models.

# 2 . The saddle-point approximation of G-L mean field theory

For a Z(N) spin system, spin variables are defined as: $s(a_i) = e^{i2\pi a_i/N}$ (lattice vertices $i = 1 \dots i_{max}$) with $a_i = 0,1,2,3 \dots N-1$. The mean value of the spin [***Cheuk-Yin Wong1994]***:

$$\varphi(\lambda) \equiv < Re\, s(a_i) >= \frac{\sum_{a_i} Re\, s(a_i) \exp\{\lambda Re\, s(a_i)\}}{\sum_{a_i} \exp\{\lambda Re\, s(a_i)\}} \quad (1)$$

From the above definition we see that $\varphi(\lambda)$ is the order parameter of spins system. Indeed, when the $< Re\, s(a_i) >$ is zero there is a symmetry with respect to a random distribution of the different orientations of $s(a_i)$ at different locations of the Lattice. When the temperature drops under its critical value $< Re\, s(a_i) > \neq 0$ because now there is a definite alignment of the spins at various locations of the lattice , and the symmetry of the system with respect to a random orientation of the spins is broken. This is the SB of the phase transition. The $\varphi(\lambda)$ is estimated explicitly for Z(2) and we obtain [***Cheuk-Yin Wong 1994***]

$$\varphi(\lambda) = \frac{e^{\lambda} - e^{-\lambda}}{e^{\lambda} + e^{-\lambda}} = tanh\lambda \quad (2)$$

From equation (2) it is obvious that when $\varphi(\lambda) = 0$ ( symmetric phase) then $\lambda = 0$ too and when $\varphi(\lambda) \neq 0$ (S.B) then $\lambda \neq 0$ . So the parameter $\lambda$ has a similar role to the parameter order $\varphi$. In parametric space $\lambda$ we consider the effective Hamiltonian

$$H[\varphi] = \int (d\lambda)[\frac{1}{2}|\nabla_{\lambda}\varphi|^2 + U(\varphi)]$$

In the framework of Landau mean field theory without external field [***Huang K 1987***] :

$$U(\varphi) = \frac{1}{2} r_o \varphi^2 + \frac{1}{4} u_o\, \varphi^4 \quad (3)$$

Where $u_o > 0$ and $r_o = a_o t = a_o \frac{T-T_c}{T_c}$. Critical configurations are produced through the saddle-point approximation and give the E-L equation of "motion" [***N. Antoniou et al 1998, 2000] :***

$$\frac{d^2\varphi}{d\lambda^2} = -\frac{\partial U(\varphi)}{\partial \varphi}$$

$$\frac{d^2\varphi}{d\lambda^2} + r_o\, \varphi + u_o \varphi^3 = 0 \quad (4)$$

# 3. The construction of map for Symmetry-Breaking phenomenon

From equation 4 we produce the two dimensional Poincare map. The reduction to discrete map will take place within the parametric space λ. On the other hand this reduction will be done via the transition n→n+1, where n the "length" of the map trajectory in the space or the time. Thus, the parameter λ must be connected with n. A way to achieve this is the inductive relations known as generator functions of the renormalization group Hu-Rudnick . Therefore we consider the linear relation between λ and n:

$$\lambda_n = kn \quad (5)$$

$$\lambda_{n+1} = kn + k$$

$$\lambda_{n+1} = \lambda_n + k \quad (6)$$

So, starting from the equation (6) we result to a Hu-Rudnick type relation. Therefore we proceed to the replacement λ→kn. Then the equation (4) can be written as :

$$\frac{d^2\varphi}{dn^2} + k^2 (r_o\, \varphi + u_o \varphi^3) = 0 \quad (7)$$

Using the relations :

$$\psi = \dot{\varphi}$$

$$\dot{\psi} = -k^2\, (r_o\, \varphi + u_o \varphi^3)$$

the equation (7) gives the following two dimensional Poincare map.

$$\varphi_{n+1} = \varphi_n + \psi_n \quad (8)$$
$$\psi_{n+1} = \psi_n - k^2 \varphi_n (r_o + u_o \varphi_n^2) \quad (9)$$

According to G-L theory when the temperature drops under its critical value, the SB phenomenon takes place. The fixed point of symmetric phase becomes unstable and its position converts to two new fixed points at symmetric positions:

$$\varphi_{1,2} = \pm\left(\sqrt{\frac{\alpha_o}{u_o}}\left|t^{1/2}\right|\right) = \pm\sqrt{\frac{|r_o|}{u_o}} \quad (10)$$

To confirm the model of the two-dimensions map (8), (9) we will reproduce the SB phenomenon in the phase space (ϕ,ψ) . The SB will appear when the temperature drops under its critical value ($r_o$ <0).

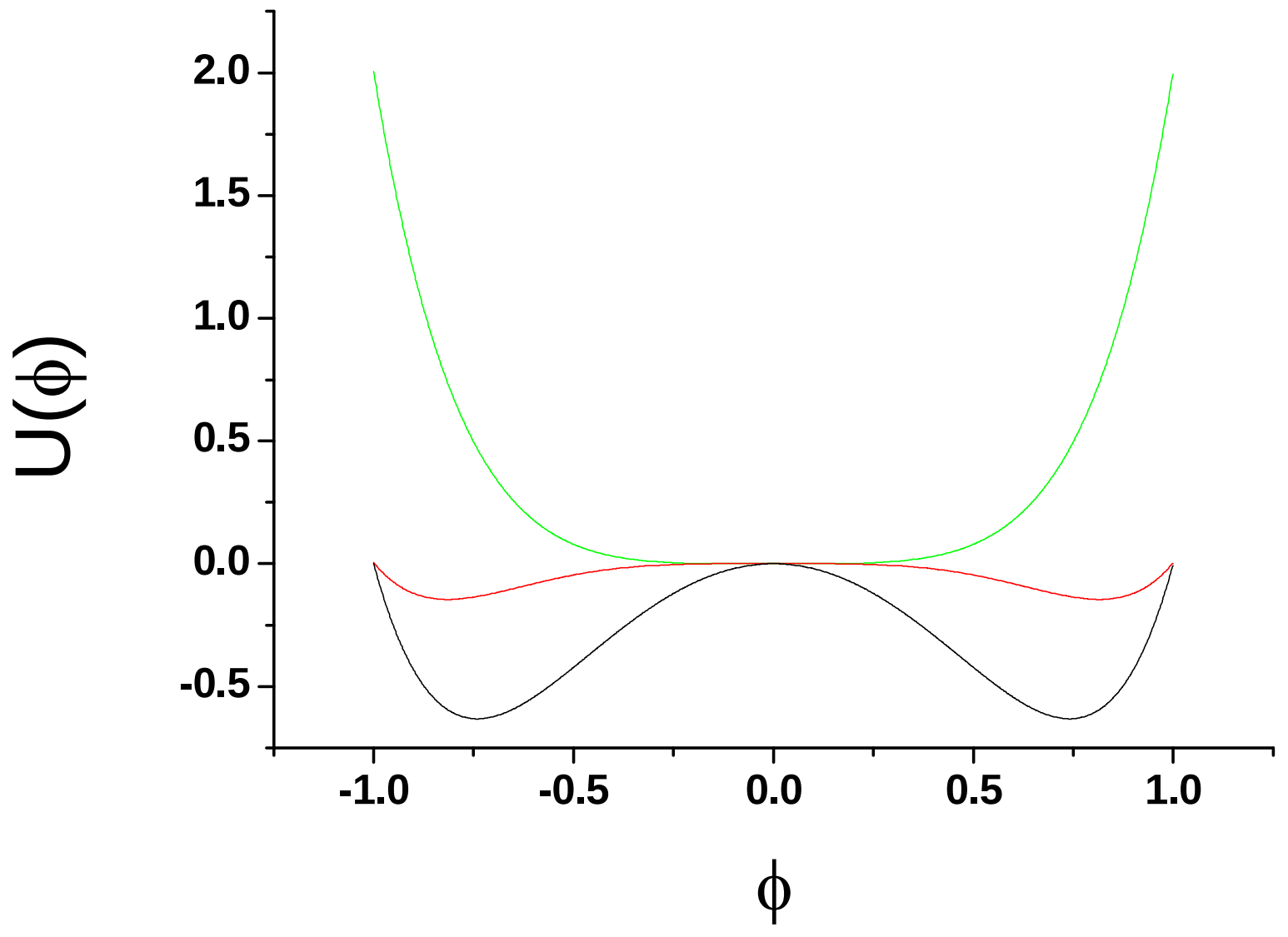

***Fig.1 The G-L free energy (3) is shown. With green line the symmetric phase*** $\boldsymbol{r_o} > 0$***. With red and black lines the broken symmetry phase*** $\boldsymbol{r_o} < 0$

We run the two dimensional map until the maximum value of iterations n becomes 100000. Due to the fact that there are is not any relation between $r_o$ and $u_o$ we can give in these parameter the same value. In figure (2) the symmetric state in phase space $(\varphi, \psi)$ is shown.

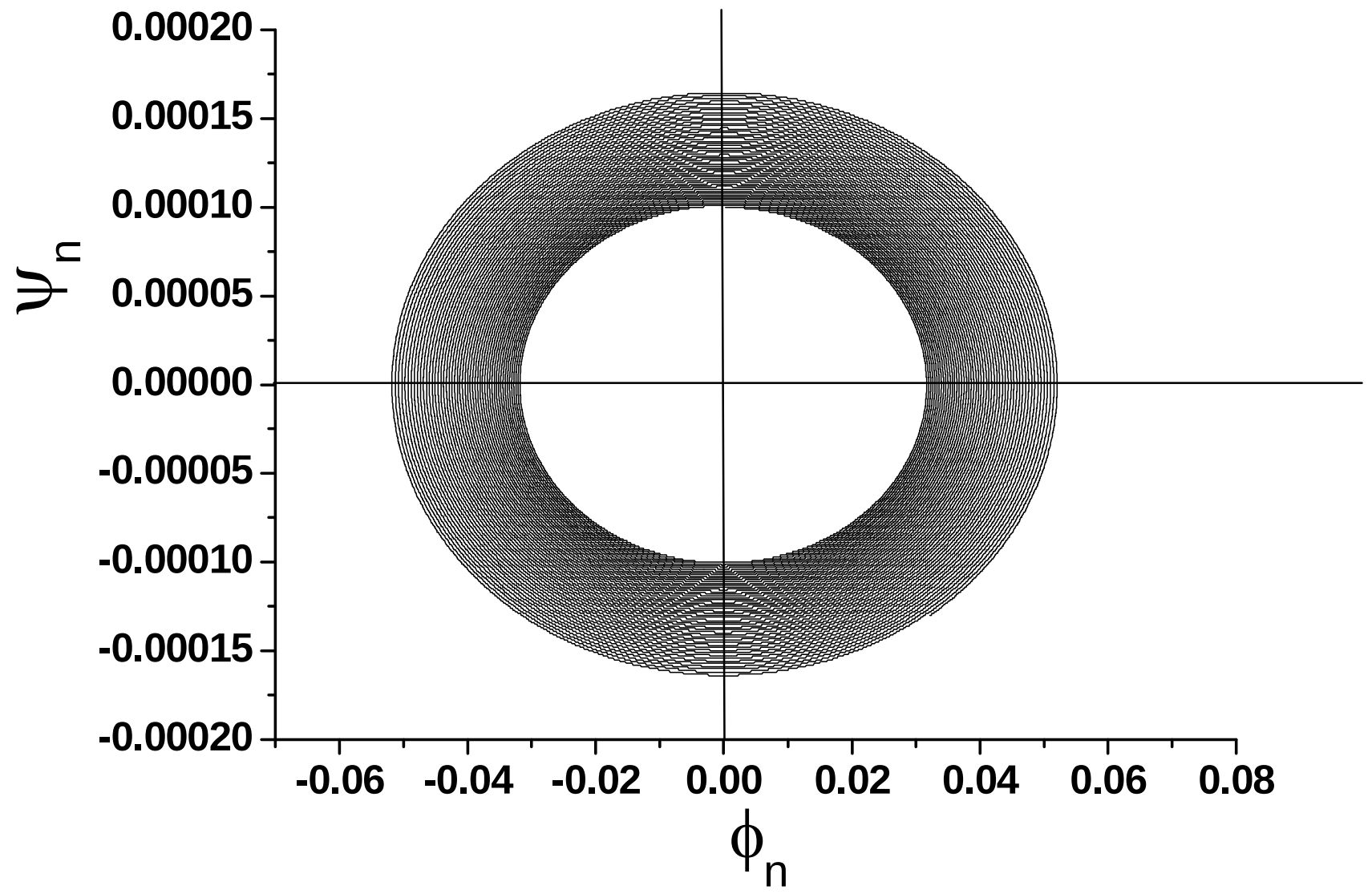

***Figure 2. The phase space diagram for symmetric state. The parameter values are: $k^2$=$10^{-5}$, $r_o = u_o = 1$, $n_{max} = 100000$. The initial values are $\varphi_1 = \psi_1 = 10^{-4}$ . There is a fixed point at the centre (0,0) as expected.***

In figure 3 the symmetry breaking for $r_o < 0$ in phase space is shown.

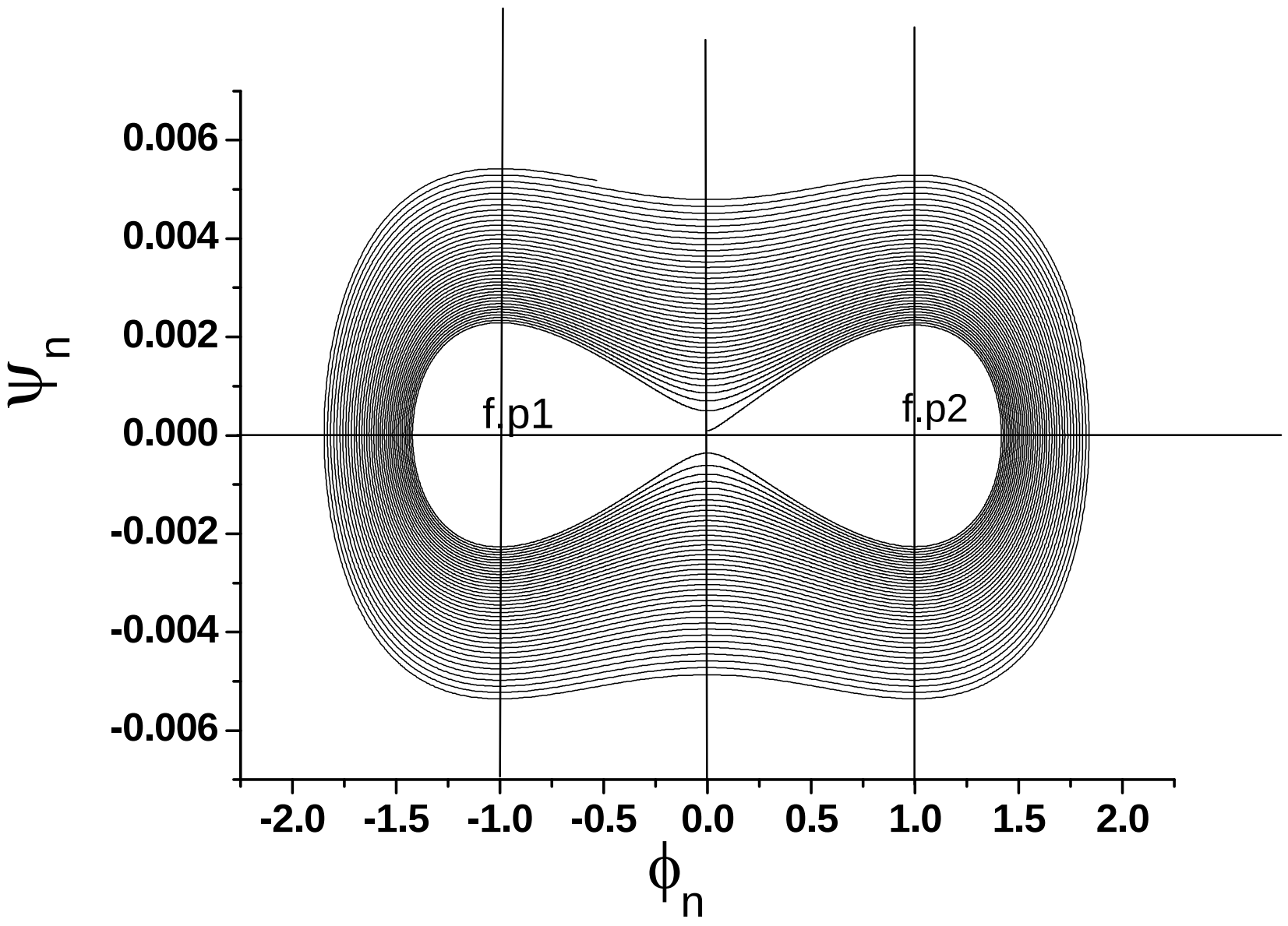

***Figure3. The phase space diagram for SB state. The parameter values are : $k^2$=$10^{-5}$, $r_o = -1$ , $u_o = 1$ , $n_{max} = 100000$. The initial values are $\varphi_1 = \psi_1 = 10^{-4}$ . There are two new fixed points at the positions (-1,0) and (1,0) as it results from equation (10). These points are not the centres of the lobes because the symmetry has been broken.***

The above diagram has a serious problem. The two lobes communicate to each other. This means that SB can not be completed. So we have to find a way to deal with this problem by making the right correction in the two dimensional map. We present such a way in section 6.

# 4 . The tricritical crossover

Beyond of the mean field theory of the G-L free energy, equation (3) is written with the addition of the $\varphi^6$ term as : [***Huang K 1987***]

$$U(\varphi) = \frac{1}{2} r_o \varphi^2 + \frac{1}{4} u_o \varphi^4 + \frac{1}{6} c_o \varphi^6 \quad (11)$$

where $c_o > 0$.

An interesting behavior is demonstrated at the region of parameter space where $r_o$, $c_o > 0$ are positive but $u_o$ changes sign and becomes negative. There is a region between the first-order phase transition line $u_0 = \frac{4}{\sqrt{3}} \sqrt{r_o c_o}$ [***Huang K 1987***] and the line $u_0 = 2\sqrt{r_o c_o}$ [***Y.F. Contoyiannis et al 2015***] where a "crossover" between second-order free energy to a first-order free energy is accomplished. In Fig.4 such a crossover between the second-order phase transition (Fig. 4a) and the first-order transition (Fig. 4d) is demonstrated. In Fig. 4b the form of G-L free energy on the line $u_0 = 2\sqrt{r_o c_o}$, where the deformation (crossover) from the second-order phase transition to the first-order transition begins, is shown. Fig. 4c depicts an intermediate state.

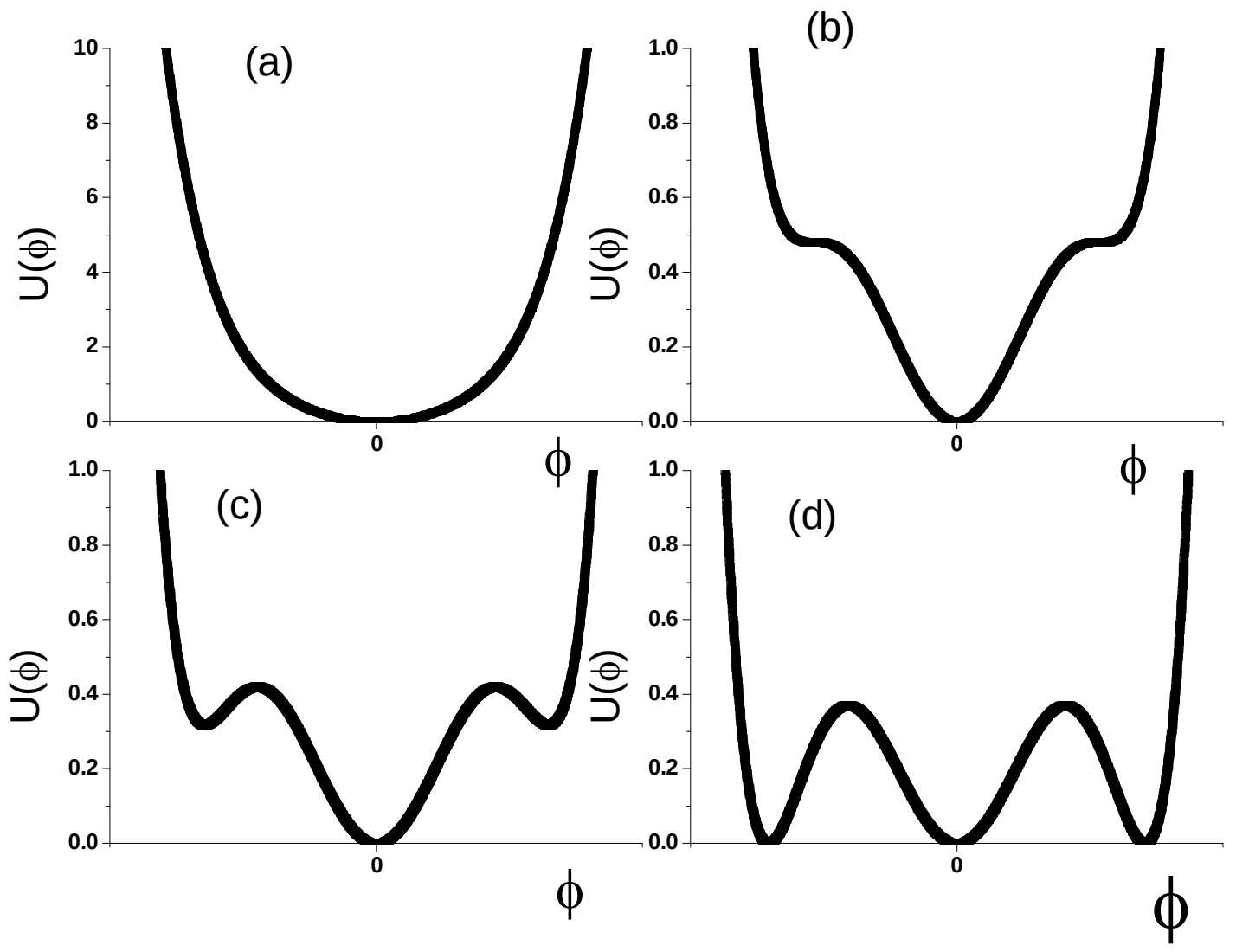

***FIG. 4 (a) The second-order phase transition free energy. (b) The beginning of energy deformation. (c) The crossover as metastable phase. (d) The first-order phase transition free energy.***

In this intermediate state the system has properties of the second-order phase transition (like power-laws) as well as of the first-order phase transition (discontinuous changes in the order parameter). Such a behavior is named as *tricritical behavior*. This behavior is a metastable phase between the second-order free energy and first-order free energy. Generally the parameter $u_o$ in the tricritical crossover is given as

$$u_0 = -h\sqrt{r_o}c_o \quad (12)$$

where the parameter h takes values in the interval $2 \leq h \leq \frac{4}{\sqrt{3}}$..

# 5. The construction of map for tricritical crossover

Following the process which we have presented in the previous sections starting from the eq.(11) we take the two dimensional map for the tricritical case:

$$\varphi_{n+1} = \varphi_n + \psi_n \quad (8)$$

$$\psi_{n+1=}\psi_n - k^2\varphi_n(r_o + u_o\varphi_n^2 + c_o\varphi_n^4) \quad (13)$$

where $u_o$ is given by eq.(12).

Letting $k^2 = g$ , the above maps take the form:

$$\varphi_{n+1} = \varphi_n + \psi_n \quad (8)$$

$$\psi_{n+1=}\psi_n + g\varphi_n(-r_o + h\sqrt{r_o}c_o\varphi_n^2 - c_o\varphi_n^4) \quad (14)$$

In fig.5 the phase space portrait of the map (8),(14) for different values of parameter h is presented.

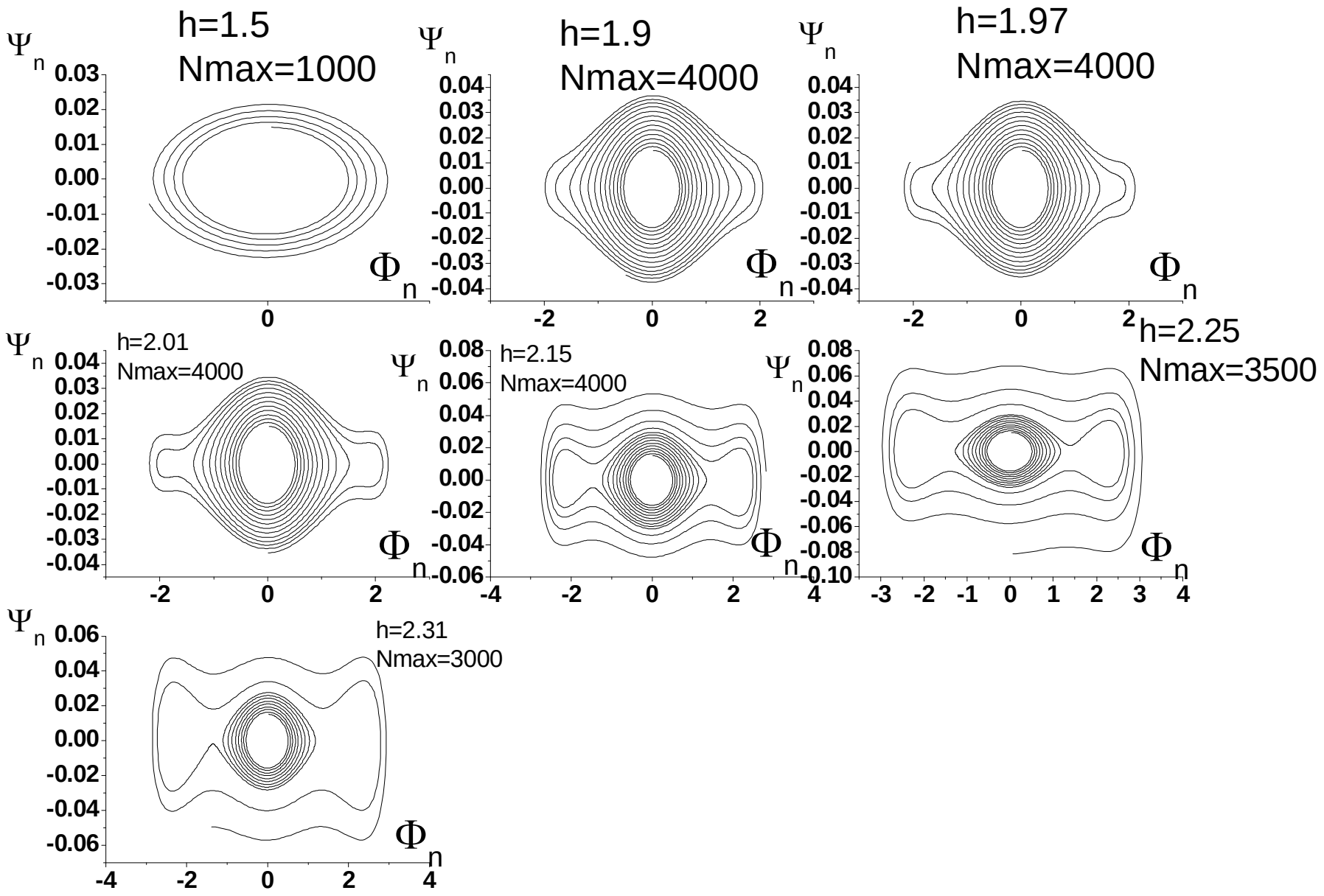

*Fig.5. The above plots have been taken for g=0.0001, $r_o = 10$, $c_o = 1$. The initial values are $\varphi_1 = 0.01, \psi_1 = 0.015$. The transition to the metastable phase between h=1.97 and h=2.01 with the appearance of three fixed points in phase space is presented.*

# 6. A correction when the order parameter $\varphi$ takes small values close to zero

Approaching the critical state, the order parameter $\varphi$ takes small values close to zero. As we see from equation (2) this means that the parameter $\lambda$ takes small values too. Therefore, we will try to find the appropriate $\lambda$ approximation so that the correction that this approach will impose on the map is capable of closing the communication of the two lobes.

We develop the order parameter $\varphi(\lambda) = tanh\lambda$ for small values $\lambda$ of keeping up to two order .

$$\varphi = tanh\lambda = \frac{sinh\lambda}{cosh\lambda} = \frac{\lambda}{1 + \frac{\lambda^2}{2!}}$$

$$= \frac{\lambda}{e^{\lambda} - \lambda} = \frac{1}{\frac{e^{\lambda}}{\lambda} - 1}$$

We consider the approach $\frac{e^{\lambda}}{\lambda} \gg 1$ where we will have that:

$$\varphi = \lambda e^{-\lambda} \quad (15)$$

or in discrete form using equation (5):

$$\varphi_n = kne^{-kn} \quad (16)$$

we are modifying the equation (8) as:

$$\psi_n = \varphi_n - \varphi_{n-1} \ (17)$$

Using equation (16) we take

$$\psi_n = ke^{-kn}(n - ne^k + e^k) \ (18)$$

From eq. (18) we estimate the quantity

$$\psi_{n+1} - \psi_n = ke^{-k(n+1)}\,[n + 1 - (n + 1)e^k + e^k]\text{-}\,ke^{-kn}(n - ne^k + e^k)$$
$$= \ ke^{-kn}[(n + 1)e^{-k} - 2n + ne^k - e^k]$$
$$= \ ke^{-kn}[n(e^k + e^{-k}) - (e^k - e^{-k}) - 2n]$$
$$= 2\ ke^{-kn}(ncoshk - sinhk - n)$$

In the framework of the approximation which we have made at the beginning of section we keep terms up to two order and so we obtain:

$$\psi_{n+1} - \psi_n = 2\ ke^{-kn}\left(n\left(1 + \frac{k^2}{2}\right) - k - n\right)$$
$$= k^3ne^{-kn} - 2k^2e^{-kn} \quad (19)$$

On the other side of equation (8) the quantity $k^2\varphi_n(r_o + u_o\varphi_n^2)$ is written as :

$$k^2\varphi_n(r_o + u_o\varphi_n^2) \ = k^2kne^{-kn}[r_o + u_ok^2n^2e^{-2kn}]$$
$$= u_ok^3\ ne^{-kn}[\tfrac{r_o}{u_o} + (\tfrac{\lambda}{e^\lambda})^2] \quad (20)$$

In SB $r_o < 0$ where the eq.(20) is written as :

$$k^2\varphi_n(-|r_o| + u_o\varphi_n^2) \ = u_ok^3\ ne^{-kn}[\tfrac{-|r_o|}{u_o} + (\tfrac{\lambda}{e^\lambda})^2] \quad (21)$$

For the reasons which we refer in section 2 we can take $|r_o| = u_o = 1$ and so the equation (21) gives that :

$$k^2\varphi_n(-|r_o| + u_o\varphi_n^2) \ = k^3\ ne^{-kn}\left[-1 + \left(\tfrac{\lambda}{e^\lambda}\right)^2\right] \quad (22)$$

According to the approximation which we have made in this section $\frac{e^\lambda}{\lambda} \gg 1$ the equation (22) is written as :

$$k^2\varphi_n(-|r_o| + u_o\varphi_n^2) \ = -k^3\ ne^{-kn} \ (23)$$

Comparing the map equation (8) with (19) and (23) we see that the corrective term is the quantity $-2k^2e^{-kn} = -2\,k\frac{\varphi_n}{n}$.

Therefore the two dimension Poincare map is written as:

$$\varphi_{n+1} = \varphi_n + \psi_{n+1} \quad (24)$$

$$\psi_{n+1} = \psi_n + k^2\varphi_n(1-\varphi_n^2) - 2\,k\frac{\varphi_n}{n}. \quad (25)$$

In figure 3 the phase –space of map (24)-(25) is shown.

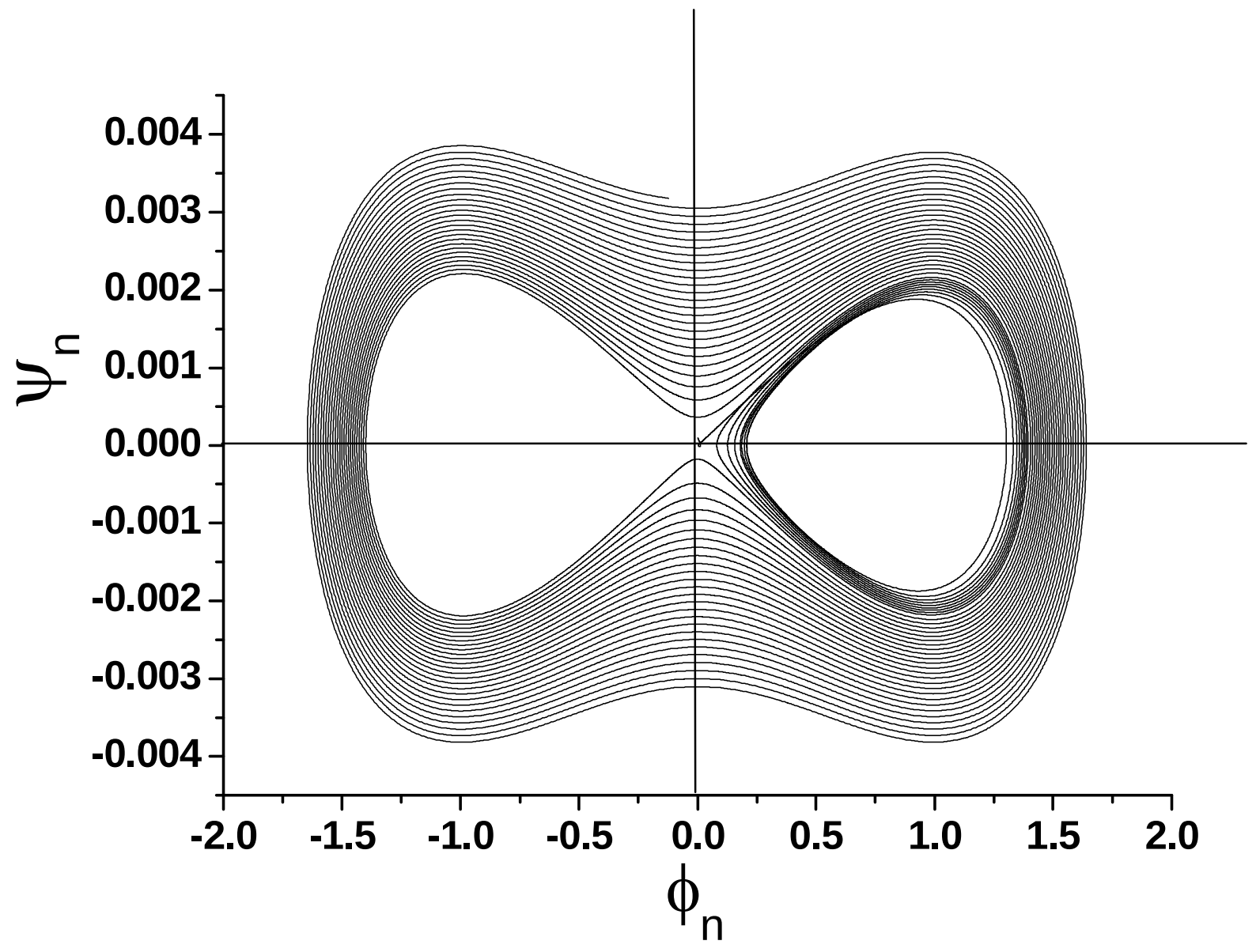

***Figure 6 . The phase space diagram for SB state with the correction. The parameter values are :*** $k^2 = 10^{-5}, n_{max} = 100000$***. The initial values are*** $\varphi_1 = \psi_1 = 10^{-4}$ ***.***

As we see from fig6 now the two lobes no longer communicate to each other. Similar behavior we take in the case of tricriticality too, if we impose the above correction. We finish this section making the notice that the equation (16) violates the causality . This is a very interesting issue which needs further investigation about its impact on the theory of critical phenomena.

# 7. Numerical models.

We will confirm the phase space diagrams produced from the corresponding maps of the SB and the tricriticality by producing corresponding phase space diagrams from numerical experiments which demonstrate these phenomena. For this scope, we will consider Z (N) spin models according to what we said at the beginning of the second section. Specifically for N = 2 and for 3 dimensions we consider the 3D-Ising model and N = 3 in 3 dimensions the Z (3) spin model. An effective algorithm which produce configurations is the Metropolis algorithm. In this algorithm the configurations at constant temperatures are selected with Boltzmann statistical weights $e^{-\beta H}$ , where H the Hamiltonian of the spin system with nearest neighbors interactions which can be written as

$$\mathrm{H} = -\sum_{<i,j>} \mathrm{J_{ij} s_i\, s_j}$$

# 7.1 The 3D-Ising model.

It is known [Huang K, 1987] that this model undergoes a second-order phase transition when the temperature falls below a critical value.

Thus for a $20^3$ lattice the critical temperature has been found to be T = 4.545 ($\mathrm{J_{ij}}$ =1) [Y. Contoyiannis 2002]. Therefore for temperatures smaller than the above value we will have a SB phenomenon. The quantity we record in the numerical experiment performed with the Metropolis algorithm is the mean magnetization M that plays the role of the order parameter. The trajectory generated by the numerical experiment is a "timeseries" of the fluctuations of the order parameter. For this trajectory we record the phase diagram ( $M_n, M_{n+1} - M_n$ ) for the critical temperature and for a temperature close but less than the critical one. The results are shown in Figure 7.

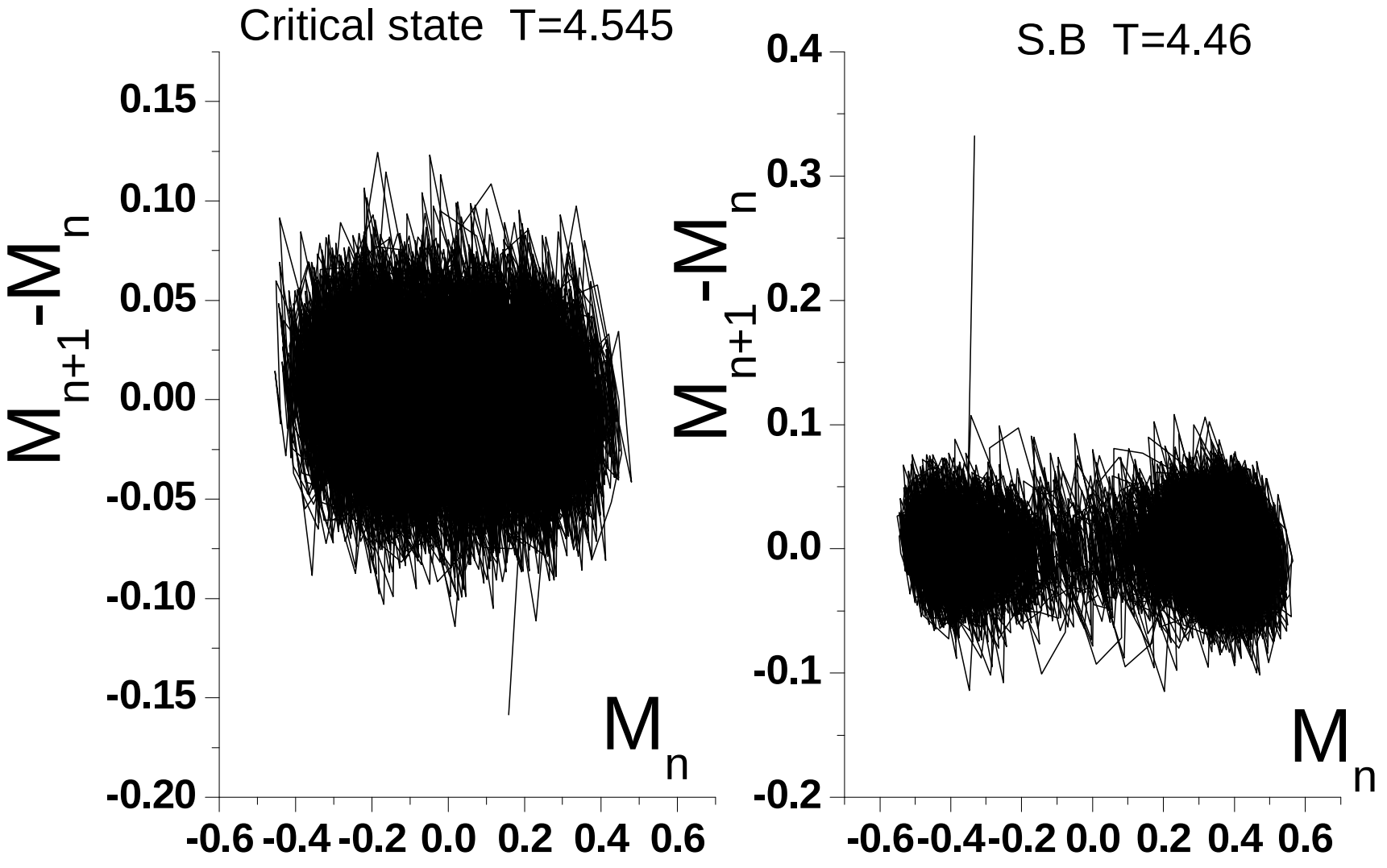

***Fig.7 (a) The one lobe of critical state is shown. (b) The two lobes of the SB are shown. The noise that appears is due to the random numbers introduced by the Metropolis algorithm.***

## *7.2 The Z(3) spin model.*

It is known [***Y.F. Contoyiannis and F.K. Diakonos 2007***] that a system which present a tricritical crossover could be the Z(3) spin system. Indeed, we found that for a lattice $20^3$ at T=2.7268 ($J_{ij}$ =1) the projection of the mean magnetization in y-axis of lattice obeys to tricritical dynamics, as in phase space of the corresponding trajectory (Fig.8).

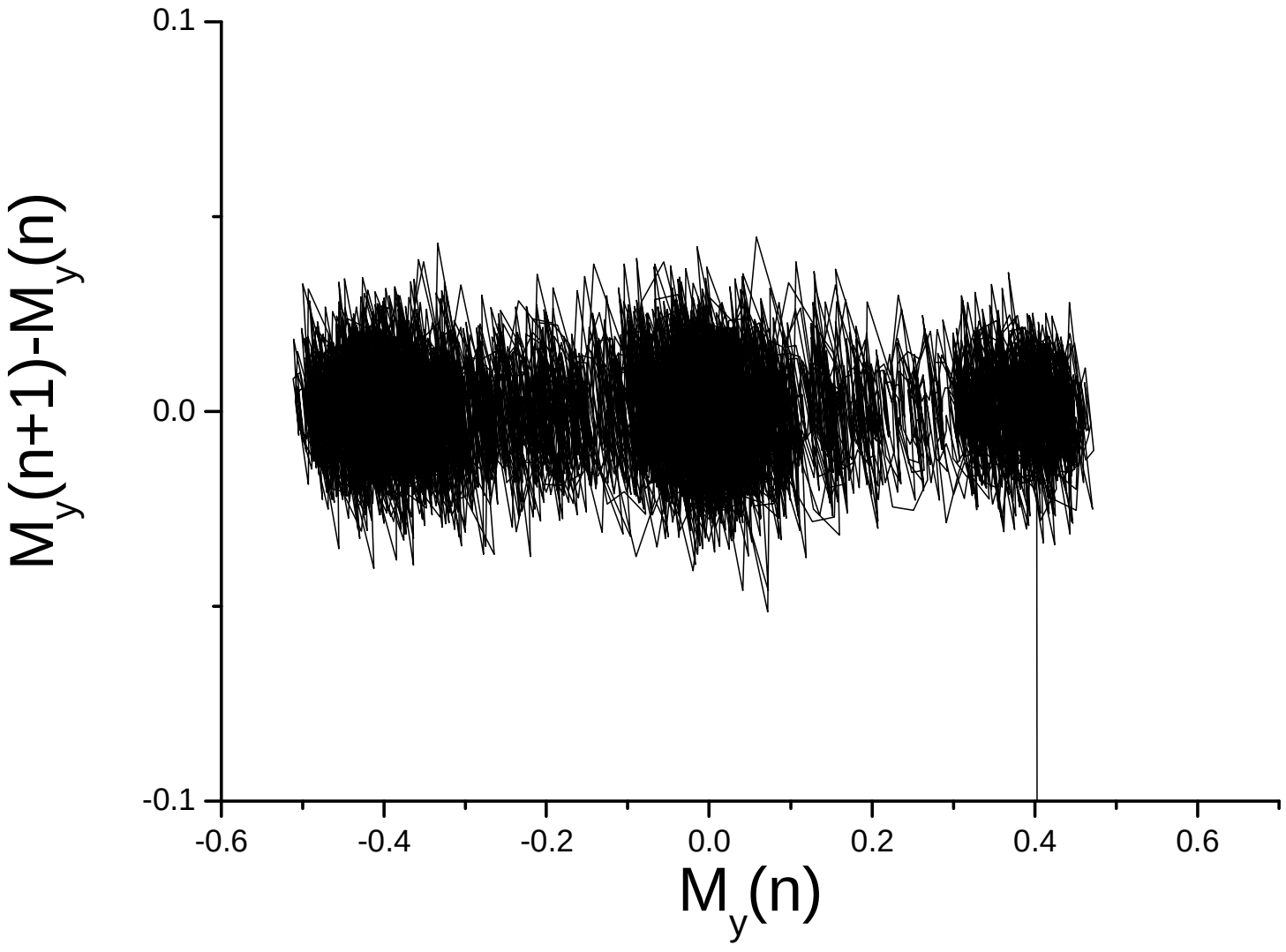

.

***Fig.8. The lattice is $20^3$ and the temperature is T=2.7268. The phase space diagram of the trajectory of the mean magnetization on y-direction of lattice is shown. The appearance of three lobes is obvious as it expected from the tricritical crossover.***

# 8. Conclusions

In this work we have produced two-dimensional maps describing SB and tricriticality in the critical phenomena. The phase space diagrams produced by these Poincare maps confirm the approximation of the theory we used. Then we presented phase space diagrams from numerical experiments of spins models, that despite the noise of the algorithms confirm the maps we have produced. In this paper, we did not get involved with the basic properties of Poincare maps that are based on the standard processes [***H.G. Schuster. 1998***] that is the calculation of Liapunov exponents, correlation functions and invariant density. These issues will be the subject of future work.